\documentstyle[a4,12pt]{article}

\begin{document}

\title{Simple classical model of spin membrane particle}
\author{E. P. Likhtman\\
All - Russian Institute for Scientific and Technical Information\\
(VINITI), Usievicha 20, Moscow 125315, Russia\\
e-mail: peisv@viniti.ru}
\maketitle

\begin{abstract}
In this paper I developed a classical model of elementary particle
that is associated with a membrane of finite size, surrounded by
non-linear electromagnetic field. The form of local interaction
which lead to bounded states of finite masses, charges and spins was
constructed. To do this I added Kaluza-Klein Lagrangian on the
surface, which is associated with extended particle, to quadratic in
field potentials term (a la tachyon mass). This ensures that
Lagrangian remains an even function of the field, but spontaneous
symmetry breaking leads to nontrivial soliton-like solutions. I
assumed that the particle has axial symmetry and that the surface has
only one degree of freedom, i.e. is a disk with the radius determined
from the equations of motion. The solution of system of two non-linear
partial differential equations for the field potentials was obtained
numerically by different methods. Several solutions with increasing 
orders of leading field multipoles and disk radius were obtained, 
and masses, electric charges and spins were calculated. In the framework 
of this model the ratio of a charge square to a double spin, i.e. 
the fine structure constant, which do not depend on parameters of the model, 
was calculated.
\end{abstract}


\section{Introduction}

Physicists handle the laws of nature in two ways: they either explain them using
more fundamental hypotheses or just postulate them. The values of electric
charges and spins of elementary particles are still almost postulated in the
framework of quantum relativistic field theory, whereas their masses became
a subject of non-trivial calculations a long time ago. Wineberg-Salam model
predicted not only existence and characteristics of intermediate bosons but
allowed to calculate electroweak processes in tree-like approach. However,
the calculation of radiation corrections is limited by the absence of
experimental data on Higgs bosons' masses. Also further unifications of
interactions such as supersymmetry~\cite{GL} and great unification~\cite{VN}
were not supported by clear experimental observations so far, and
electrodynamics are still leading in calculation precision of stationary
states. Nevertheless electrodynamics can not be considered as an ideal model
for the lepton world because of two problems: divergences in both classical
and quantum theories, and the absence of calculations of some dimensionless
quantities such as fine structure constant and lepton mass ratios.

Non-linear extensions of electrodynamics were first proposed by 
Kaluza~\cite{KK1}, Klein~\cite{KK2} as well as by Born and Infeld~\cite{BI}. Non-linearity%
leads to non-singular solution in case of point-like particle~\cite{RK}, but
the particle mass, charge and magnetic moment remain arbitrary. To change
this situation one can introduce a fundamental length into the model while
preserving the local character of interaction, and construct a non-singular
solution with finite size field sources.

In the present paper I propose an extended particle model, which consists of
only one real vector electromagnetic field. The model is based on bilinear
Maxwell Lagrangian
\begin{equation}
L_M=\frac{1}{8\,\pi}\,\int({\it E}^{2}-{\it B}^{2})\,dV,  \label{lm}
\end{equation}
and quadrolinear Kaluza-Klein Lagrangian (self-interaction of vector field)
\begin{equation}
L_{K-K}=\frac{1}{8\,\pi \,\beta ^2}\,\int {({\it E\cdot B}})^{2}\,dV,
\label{lkk}
\end{equation}
where constant $\beta $ has dimension of field strength.

The key assumption of this work is proportionality of the four-current on
some surface associated with the particle to the four-potential of the
field. Therefore the gauge invariance is broken on the surface. In order to
generate this dependence in the form of Lagrange equations I add  quadratic
in potentials term on the surface to the Lagrangian
\begin{equation}
L_{surface}=-\frac{1}{4\,\pi \,l}\,\int ({\it p}^{2}-{\it A}^{2})\,dS,
\label{ls}
\end{equation}
where $l$ is a fundamental length and $p$ and $A$ are scalar and vector 
potentials. The Lagrangian remains an even function
of the field, and to provoke spontaneous symmetry breaking the sign of the
last term must be the same as the sign used in the theory of vector tachyon
field.

The solution of the problem was found by using approximation of infinite
number of degrees of freedom by the finite number. It appears that
non-linearity of Kaluza-Klein interaction on the one hand and the
competition of surface and volume energy on the other hand lead to a
spectrum of non-trivial stationary states with the finite values of all
observables. This result was obtained by assuming an axial symmetry of the
whole system, and that the surface is a disk with only one degree of freedom
--- disk radius. The description of the field by only two degrees of freedom
has already led to a set of stationary solutions. Then more field degrees of
freedom were added (up to 1000) so that solutions achieved an asymptotic
behavior.

An important feature of the presented classical model is that the charges,
magnetic moments and spins are no longer parameters of the Lagrangian, but
arise due to a spontaneous symmetry breaking. These observables are
proportional to some powers of two fundamental parameters of the Lagrangian
$\beta$ and $l$. To compare predictions of the model with experiment one can
construct several dimensionless parameters, which do not depend on $\beta$
and $l$. Therefore, in the following text I set these two parameters to
unity without loss of generality. For each state I calculate Lande
coefficient and the ratio of a charge square to a double spin $e^2/(2\,s)$,
i.e. the fine structure constant. I also extract the ratios of masses, spins
and magnetic moments of different states.

\section{Lagrangian in ellipsoidal coordinates}

Owing to an axial symmetry and particular choice of calibration the problem
was reduced to solution of two non-linear partial differential equations for
electric field potential $p$ and the only non-zero component of
vector-potential $A=A_{\phi}$, which both depend on two spacial
coordinates. I introduced Cartesian coordinates so that $z$ is perpendicular
to the disk plane, and then did the following transformation to the
ortogonal ellipsoidal coordinates
\begin{equation}
x=\rho \,\cos \left( \phi \right),\label{xx}
\end{equation}
\begin{equation}
y=\rho \,\sin \left( \phi \right),\label{yy}
\end{equation}
\begin{equation}
z=R\,\frac{\bar v(v)\,\cos \left( \pi/2\,\bar w(w)\right) }{\sin
\left( \pi/2
\,\bar w(w)\right) },\label{zz}
\end{equation}
\[
\rho ={R\,\frac{\sqrt{1-\bar v(v)^{2}}}{\sin \left( \pi/2\,\bar
w(w)\right) }},
\]
where $R$ is the disk radius. This transformation reduced an infinite space
to a unit square ($w,v$) (note that none of the functions depends on the
third angular coordinate $\phi $). I have concentrated on the mirror
symmetry case $L(-v)=L(v)$, and therefore evaluated all functions in the 
interval $0\leq v\leq 1$. The transformation \ref{xx}--\ref{zz} can be 
tuned by two monotonically increasing functions $\bar w(w)$ and $\bar v(v)$ 
($0\leq w,\bar w,\bar v\leq 1$). The simplest case is 
$\bar w(w)=w,\ \bar v(v)=v$, which we explore in subsection~\ref{basis}. 
The particular choice of functions $\bar w(w)$ and $%
\bar v(v)$ should not affect the values of observables, but could be used to
achieve better convergence to the asymptotic solution. I have also
introduced contravariant potentials
\begin{equation}
\tilde{\it p}=p\,R,  \tilde{\it A}=A\,R\,h_{\phi},\label{Ap}
\end{equation}
where $h_{\phi }=\sqrt{g_{\phi \phi }}/R$ is dimensionless Lame coefficient,
and ''field strength''
\begin{equation}
E_{w}=-{\frac{d}{dw}}\tilde{{\it p}}, E_{v}=-{\frac{d}{dv}}\tilde{{\it
p}},\label{E}
\end{equation}
\begin{equation}
B_{w}={\frac{d}{dv}}\tilde{{\it A}}, B_{v}=-{\frac{d}{dw}}\tilde{{\it A}}.
\label{z}
\end{equation}

The boundary conditions on the axis $v=1$ are $\tilde{{\it A}}=0$ with
finite $E_v$, and at infinity $w=0$ are $\tilde{{\it p}}=0$, $\tilde{{\it
A}}%
=0$. The boundary conditions in the disk plane outside the disk itself $v=0$
are $E_v=0$ and $B_w=0$. Now we can write Lagrangian using new variables as
\begin{equation}
L=L_p+L_A+L_{K-K}, L_p=L_{pv}+L_{ps}, L_A=L_{Av}+L_{As},\label{L}
\end{equation}
where subscripts $v$ and $s$ denote volume and surface terms, and
\begin{equation}
L_{pv}=-\frac {R}{2}\,\int _{0}^{1}\!\int
_{0}^{1}\!(F^w\,E_w^2+F^v\,E_v^2)dv\,dw,
\label{Lp}
\end{equation}
\begin{equation}
L_{Av}=\frac {R}{2}\,\int _{0}^{1}\!\int
_{0}^{1}\!(G^w\,B_w^2+G^v\,B_v^2)dv\,dw,
\label{LA}
\end{equation}
\begin{equation}
L_{K-K}=\frac {1}{2\,R}\,\int _{0}^{1}\!\int _{0}^{1}\!K\,(B\cdot
E)^{2}\,dv\,dw,
\label{LKK}
\end{equation}
\begin{equation}
L_{ps}=\frac {R^2}{2}\,\int _{0}^{1}\!S^p\,\tilde{{\it p}}^2\,dv,
\label{Lps}
\end{equation}
\begin{equation}
L_{As}=-\frac {R^2}{2}\,\int _{0}^{1}\!S^A\,\tilde{{\it A}}^2\,dv,
\label{LAs}
\end{equation}
where
\[
F^w=h_v\,h_\phi/h_w, F^v=h_w\,h_\phi/h_v,
\]
\[
G^w=1/F^w, G^v=1/F^v,
\]
\[
K=1/(h_v\,h_w\,h_\phi), (B\cdot E)=(B_w\,E_w+ E_v\,B_v)
\]
\[
S^p=h_v\,h_\phi, S^A=h_v/h_\phi.
\]

Orbital contribution to the angular momentum is zero due to the axial
symmetry $E_{\phi }=0$. The $z$-component of the spin is determined by the
integral
\begin{equation}
s=R^2\,\int _{0}^{1}\!\int _0^1\!(  D_w\,\frac {dw}{d\rho}\,h_w^2+  D_v\,%
\frac {dv}{d\rho}\,h_v^2 )\, \tilde{{\it A}}\,h_\phi\,dv\,dw,
\label{s}
\end{equation}
where
\[
D_w= E_w+ B_w\,(B\cdot E)\,K/F^w,D_v= E_v+ B_v\,(B\cdot E)\,K/F^v.
\]

\section{Numerical solution}

In order to solve the model numerically, I approximated the fields by
the finite number of degrees of freedom either by expansion of both
potentials in series or by discretization of the unit square. The third
method combines both approaches. In all methods the problem was reduced to a
solution of the system of many non-linear algebraic equations. The system
was solved by the iterative Newton's tangent method for many variables. I
supposed that the approximate asymptotic solution was found if solutions
converged and the energy values did not change significantly with increasing
number of variables. To choose initial approximation I have used the
property of Kaluza-Klein interaction (generally speaking, of any Lagrangian
with only quadratic and fourth-order dependence on the fields), that
\begin{equation}
L_{A}+L_{K-K}=0,L_{p}+L_{K-K}=0,
\label{LL}
\end{equation}
on exact solutions. To ensure positive values of energy $H$ (in the
following I use $H=-L$) one requires $H_{A}=H_{p}=-H_{K-K}>0$.

\subsection{Basic functions method}

\label{basis}

The simplest method, which allows to find qualitative solutions with only
two field degrees of freedom, is basic functions method. In this method one
expresses potentials as a finite series of normalized functions with unknown
amplitudes, so that Lagrangian becomes a function of these amplitudes.
Calculation of bilinear part of Lagrangian requires evaluation of relatively
small number of integrals, whereas number of integrals required for
quadrolinear terms is proportional to fourth power of basic functions
number. Therefore the advantages of the method can be seen only for the
small number of correctly chosen functions.

The natural choice of basic functions is solutions of Maxwell equations
(fixing the choice of coordinate system so that $\bar w(w)=w$ and $\bar
v(v)=v$)
with the first few multipole moments

\begin{equation}
\tilde p_1={\frac {\pi\,w}{2}},
\label{p1}
\end{equation}
\begin{equation}
\tilde A_2={\frac {\left( -\pi\,w+\sin \left( \pi \,w \right) \right) \left(
1-{v}^{2} \right) }{-1+\cos \left( \pi \,w \right) }},
\label{A2}
\end{equation}
\begin{equation}
\tilde p_3={\frac {\left( \left( 4+2\,\cos \left( \pi \,w \right) \right)
\pi \,w-6\,\sin \left( \pi \,w \right) \right) \left( 3/2\, {v}^{2}-1/2
\right) }{-1+\cos \left( \pi \,w \right) }},
\label{p3}
\end{equation}
\[
.............,
\]
which were normalized at $v=0$ and $w=1$ to be $\pi/2$. The potentials
become
\begin{equation}
\tilde p(v,w) =\sum _{k=1}\tilde p_{2\,k-1}(v,w)\, a_{2\,k-1},
\label{pvw}
\end{equation}
\begin{equation}
\tilde A(v,w) =\sum _{k=1}\tilde A_{2\,k }(v,w)\, a_{2\,k },
\label{Avw}
\end{equation}
where $a_1=e/R$, $a_2=\mu /R^2\,3/2$, $a_3$,... are unknown multipole amplitudes,
$e$ -- charge, and $\mu$ -- magnetic moment.

Integration of bilinear part of energy leads to (for the first six
multipoles):

\begin{equation}
H_{pv}=-R/2\,(\pi/2) \,(a_1^2 +16/5\,a_3^2+4096/729\,a_5^2),
\label{Hpv}
\end{equation}
\begin{equation}
H_{Av}=R/2\,(\pi/2)\,(4/3\,a_2^2+256/63\,a_4^2+16384/2475\,a_6^2),
\label{HAv}
\end{equation}
\begin{equation}
H_{ps}=R^2/2\,(\pi/2)^2\,(1/2\, a_1^2+1/2\, a_3^2+1/2\, a_5^2
              -1/2\,a_1\,a_3 - 1/9\,a_1\,a_5 -13/36\,a_3\,a_5),
\label{Hps}
\end{equation}
\begin{equation}
H_{As}=-R^2/2\,(\pi/2)^2\,(1/4\,a_2^2+11/24\,a_4^2+29/60\,a_6^2
                           -1/3\,a_2\,a_4-1/12\,a_2\,a_6-1/3\,a_4\,a_6).
\label{HAs}
\end{equation}

Integration of quadrolinear part of energy is evaluated analytically over
$v$
and numerically over $w$. An approximate values of first $3\times 3=9$
integrals are listed below
\[
H_{K-K}=1/(2\,R)\,(  2.13\,a_1^2\,a_2^2  +6.19\,a_1^2\,a_2 \,a_4
+15.4\,a_1^2\,a_4^2  +22.2\,a_3^2\,a_2^2
\]
\begin{equation}
+16.7\,a_1 \,a_2 \,a_3\,a_4  + 121\,a_3^2\,a_4^2  -12.0\,a_1 \,a_2^2\,a_3
+15.0\,a_3 \,a_4^2\,a_1  -98.5\,a_3^2\,a_4 \,a_2).
\label{HKK}
\end{equation}

To find a simplest solution with only one electric and one magnetic
multipoles, we note that the bilinear parts of the energies $H_{p}$ and $%
H_{A}$ must be positive, which is achieved if
\begin{eqnarray}
1.27 &<&R_{1}, R_{2}<3.40  \nonumber  \label{r1r10} \\
4.07 &<&R_{3}, R_{4}<5.64  \nonumber \\
7.15 &<&R_{5}, R_{6}<8.72 \\
10.3 &<&R_{7}, R_{8}<11.8  \nonumber \\
13.4 &<&R_{9}, R_{10}<15.0  \nonumber
\end{eqnarray}
where the subscripts denote the number of corresponding multipole. I
considered values of radii which satisfy each line of eq.~\ref{r1r10}
separately in order to obtain solutions with minimal energy. First pair of
multipoles is centrosymmetric electric field and dipolar magnetic moment.
Therefore the energy $H_{1-2}$ is a function of three variables: two
multipole amplitudes $a_1$, $a_2$ and the radius $R$. From 
eqs.~\ref{Hpv}--\ref{HKK} one gets:

\begin{equation}
H_{1,2}=-R/2\,(1.57\,a_1^2-2.09\,a_2^2)+ R^2/2\,(1.23\,a_1^2-0.617\,a_2^2)
-2.13\,a_1^2\,a_2^2/(2\,R).
\label{H12}
\end{equation}

The values of these unknowns are determined from solution of the system of
three algebraic equations, which are obtained from requiring that partial
derivatives of the energy in respect to these variables are equal to zero.
To obtain the next state I chose two other degrees of freedom, which are
quadrupole electric and octupole magnetic moments, and then solve analogous
equations for $H_{3,4}$, etc.

Then the obtained solutions were used as a starting point for solution of
Lagrange equations with 8 degrees of freedom in order to increase the
precision. This procedure did not lead to a convergent process for the first
pair of multipoles but did improve calculations for all others. The possible
reason is that the determinant of the matrix of linearized equations does
not have constant sign and can be close to zero, which lead to poor
convergence. The physical reason of this problem is the areas with very
strong electrical field where equation for magnetic field becomes hyperbolic
rather than elliptic.

The results of calculation for the first three pairs of leading multipoles
are listed in Table 1 ($m=H$).

\begin{center}
\begin{tabular}{|c|c|c|c|c|}
\hline
Leading    &Included& $R$  & $m$   & $e$   \\
multipoles & multipoles &      &       &       \\ \hline
1,2        & 1..2       & 2.81 & 3.57  & 3.25  \\
1,2        & 1..6       & -    & -     & -     \\ \hline
3,4        & 3..4       & 5.03 & 0.431 & -     \\
3,4        & 1..8       & 5.49 & 0.366 & 1.14  \\ \hline
5,6        & 5..6       & 8.05 & 0.337 & -     \\
5,6        & 1..10      & 8.05 & 0.923 & 2.05  \\ \hline
\end{tabular}
\end{center}

\begin{center}
{\bf Table 1}\\
\end{center}

The table shows that as expected from eqs.~\ref{r1r10} the disk size
increases with increasing number of leading multipoles. Further increase of
number of multipoles must be accompanied by simultaneous addition of new
basic functions with non-Maxwellian $w$-dependence for lower multipoles,
which would lead to significant increase of number of integrals and
calculation time.

\subsection{Lattice approach}

An alternative method is to define the potentials on some lattice and to
assume linear interpolation between lattice points. Due to local character
of interactions, presence of first derivatives only and linear interpolation
assumption the energy depends only on the values of potentials at the
neighboring sites, so that each algebraic equation contains limited number
of variables (not more than 18), which does not increase with increasing
number of lattice points. Therefore calculation time increases slower than
the cube of number of points.

Linear interpolation and integration by trapezium rule works well only for
slowly varying functions. To increase smoothness of potentials I used
specific choice of function $\bar v(v)$, which ''stretched'' oscillations of
the
potentials near the disk axis. In particular I used $\bar
v(v)=3/2\,v-1/2\,v^{3}$
and $\bar w(w)=w$. The results of calculation for different number of
lattice points are listed in Table 2.\newline

\begin{center}
\begin{tabular}{|c|c|c|c|c|}
\hline
Leading   &     Lattice        & $R$  & $m$  & $e$  \\
multipoles&      size          &      &      &      \\
\hline
3,4      &    $10\times 10$   & 6.01 & 1.40 & 3.55  \\
3,4      &    $12\times 12$   & 5.98 & 1.43 & 3.68  \\
3,4      &    $15\times 15$   & 5.96 & 1.47 & 3.81  \\
3,4      &    $20\times 20$   & 5.96 & 1.52 & 3.91  \\
3,4      &$\infty\times\infty$& 5.96 & 1.60 & 4.04  \\
\hline
5,6      &    $12\times 12$   & 9.88 & 1.83 & 3.92  \\
5,6      &    $14\times 14$   & 9.62 & 1.58 & 3.72  \\
5,6      &    $18\times 18$   & 9.38 & 1.41 & 3.69  \\
5,6      &    $24\times 24$   & 9.23 & 1.32 & 3.68  \\
5,6      &$\infty\times\infty$& 9.05 & 1.22 & 3.68  \\
\hline
7,8      &    $12\times 12$   & 15.0 & 3.61 & 5.89  \\
7,8      &    $14\times 14$   & 14.2 & 2.44 & 4.40  \\
7,8      &    $18\times 18$   & 13.4 & 1.70 & 3.96  \\
7,8      &    $24\times 24$   & 12.8 & 1.40 & 3.90  \\
7,8      &$\infty\times\infty$& 12.1 & 1.20 & 3.88  \\
\hline
\end{tabular}
\end{center}

\begin{center}
{\bf Table 2}\\
\end{center}

I used solutions of the previous section as a starting point for the lattice
method. At the end of each part of the table extrapolated values for the
infinite lattice are listed. For leading multipoles 1,2 solutions
was found for lattice 6$\times 6$ only. For leading multipoles 3,4 stable
solutions exist for a large range of number of degrees of freedom, and
observables quickly achieve their asymptotic values. Convergence gets worse
with increasing order of leading multipole because higher order multipoles
have more oscillations on the disk surface. In particular potentials of 3,4
multipole have two extremums, 5,6 --- 3 extremums and 7,8 --- 4 extremums.

\subsection{Combined method}

To improve precision of calculations and the consistency of two previous
methods I divided space into two parts: for large distances from the disk
($%
w=0..3/4$) I used basic functions method with $2\times 5=10$ functions, and
in the vicinity of the disk ($w=3/4..1$) I used lattice approach. The
results of combined method are listed in Table 3.

\begin{center}
\begin{tabular}{|c|c|c|c|c|}
\hline
Leading   &   Lattice    & $R$  & $m$  & $e$   \\
multipoles&    size      &      &      &       \\
\hline
3,4 &$   12 \times 3    $& 5.97 & 1.41 & 3.63  \\
3,4 &$   16 \times 4    $& 5.95 & 1.47 & 3.77  \\
3,4 &$   20 \times 5    $& 5.96 & 1.51 & 3.88  \\
3,4 &$   28 \times 7    $& 5.96 & 1.55 & 3.92  \\
3,4 &$\infty\times\infty /4$& 5.97 & 1.57 & 3.94  \\
    &                   &       &      &      \\
3,4 &$\infty\times\infty$& 5.96 & 1.60 & 4.04  \\
\hline
5,6 &$    16\times  4   $& 9.31 & 1.37 & 3.70  \\
5,6 &$    20\times  5   $& 9.36 & 1.49 & 3.91  \\
5,6 &$    28\times  7   $& 9.21 & 1.35 & 3.77  \\
5,6 &$    32\times  8   $& 9.15 & 1.27 & 3.65  \\
5,6 &$    40\times 10   $& 9.13 & 1.25 & 3.63  \\
5,6 &$\infty\times\infty /4$& 9.06 & 1.19 & 3.56  \\
    &                   &       &      &      \\
5,6 &$\infty\times\infty$& 9.05 & 1.22 & 3.68  \\
\hline
7,8 &$    16\times  4   $& 13.7 & 1.96 & 4.13  \\
7,8 &$    20\times  5   $& 13.1 & 1.55 & 3.87  \\
7,8 &$    28\times  7   $& 12.6 & 1.32 & 3.92  \\
7,8 &$    32\times  8   $& 12.5 & 1.27 & 3.91  \\
7,8 &$    40\times 10   $& 12.4 & 1.21 & 3.90  \\
7,8 &$\infty\times\infty /4$& 12.1 & 1.10 & 3.90  \\
    &                   &       &      &      \\
7,8 &$\infty\times\infty$& 12.1 & 1.20 & 3.88  \\
\hline
\end{tabular}
\end{center}

\begin{center}
{\bf Table 3}\\
\end{center}

The last lines of each part of this table repeats the asymptotic results of
the
previous section. The results of lattice and combined methods are in good
agreement with each other. However, combined method is 10 times faster
and therefore gives more precise results. The results for
$\bar v=2\,v-v^2$ and $40\times 10$ lattice
differ from Table 3 only by a few percent even for 7,8 multipole.

\section{Conclusions}

I proposed a new concept of extended particle that is associated with a
membrane of finite size. Charges and currents on this surface are chosen to
be proportional to the corresponding potentials of the non-linear
Kaluza-Klein electromagnetic field, so that the gauge invariance is broken
on this surface. Then I assumed the axial and mirror symmetry of the system
and existence of the only one surface degree of freedom: radius of the disk.
Number of field degrees of freedom was also chosen to be finite. Described
set of assumptions allowed me to develop a technique for iterative
calculation of the first three stationary states of the model. Then I
demonstrated that increasing number of degrees of freedom (up to 1000) leads
to the asymptotic solutions of Lagrange equations. Several states with
increasing order of leading electric and magnetic multipoles and disk radii
were constructed numerically using this method. Each state can be
potentially associated with an elementary particle.

The key result of the paper is that the values of all observables are
finite. The masses were found to decrease weakly with increasing order of
leading multipoles. The charges, magnetic moments and spins were also
calculated from the model rather than postulated. It was not obvious that
charges and spins of different states are identical or that their ratios are
rational numbers. Calculations have shown that they are equal for the first
three states within 10-20\%:
\newline
\begin{center}
\begin{tabular}{|c|c|c|c|c|c|c|}
\hline\hline
Leading   & $R$  & $m$  & $e$  & $s$  &$1/\alpha$& $\mu/\mu_B$   \\
multipoles&      &      &      &      &          &               \\
\hline
3,4       & 6.0  & 1.6  & 4.0  &  12   & 1.4      &  1.0         \\
5,6       & 9.1  & 1.2  & 3.6  &  11   & 1.5      &  1.3         \\
7,8       & 12.1 & 1.1  & 3.9  &  12   & 1.5      &  1.1         \\
\hline
\end{tabular}
\end{center}

\begin{center}
{\bf Table 4}\\
\end{center}

More detailed and extensive calculations are needed to make more precise
conclusion. From the other side, the question about equality of charges
and spins can be resolved by discovering some hidden symmetry of the
described or modified interaction. It is also possible to change
non-linearity to Born-Infeld type without adding any new parameters
to the model.

Masses, charges, magnetic moments and spins are proportional to some powers
of the two fundamental constants of the model with dimensions of length and
field strength. However, there are two observables which do not depend on
these constants --- Lande coefficient ($\mu/\mu_B$) and the ratio of charge
square to a double spin, which is equal to the fine structure constant
if one assumes that the spin is $\hbar /2$. Lande coefficient was found
about 1, and the fine
structure constant --- more than 100 times larger than expected.
Therefore, calculated values of these parameters do not allow to associate
discovered stationary states with charged leptons. Another interpretations
(quarks) would require development of a calculation technique for bound
states probably with non-additive charges and spins.

The presented method allows to modify calculations for different
non-linearity and/or boundary conditions on the surface without adding
new model parameters. This will preserve the predictive ability of
the approach and the existence of finite values of the observables.

\section*{Acknowledgment}

I thank I.V.~Tutin, B.L.~Voronov and A.E.~Likhtman for continuous support
and
discussions.

\end{document}